\documentclass[floatfix,notitlepage,prl,reprint,superscriptaddress,twocolumn]{revtex4-1}
\usepackage{graphicx,amsmath,bm,natbib,color}

\begin{document}

\title{Observing dynamical friction in galaxy clusters}
\author{Susmita Adhikari}
\affiliation{Department of Astronomy, University of Illinois at Urbana-Champaign, 1002 W. Green St., Urbana IL 61801, USA}
\author{Neal Dalal}
\affiliation{Department of Astronomy, University of Illinois at Urbana-Champaign, 1002 W. Green St., Urbana IL 61801, USA}
\author{Joseph Clampitt}
\affiliation{Department of Physics and Astronomy, Center for Particle Cosmology, University of Pennsylvania,
 209 S. 33rd St., Philadelphia, PA 19104, USA}

\begin{abstract}
We present a novel method to detect the effects of dynamical friction in observed galaxy clusters.  Following accretion into clusters, massive satellite galaxies will backsplash to systematically smaller radii than less massive satellites, an effect that may be detected by stacking the number density profiles of galaxies around clusters.  We show that this effect may be understood using a simple toy model which reproduces the trends with halo properties observed in simulations.  We search for this effect using SDSS redMaPPer clusters with richness $10<\lambda<20$, and find that bright ($M_i<-21.5$)  satellites have smaller splashback radii than fainter ($M_i>-20$) satellites at $99\%$ confidence.
\end{abstract}

\maketitle

In the standard cosmological model, cosmic structures form hierarchically.  Short wavelength perturbations collapse to generate low-mass halos, which subsequently merge to create larger halos, which themselves merge to form even larger structures, and so on.  Every macroscopic structure in the dark matter field is believed to have assembled out of constituent structure on smaller scales.  Much of that substructure survives within halos, however a considerable amount of substructure is dynamically erased, through processes like tidal stripping and dynamical friction.  In this paper, we focus on the latter process.

Dynamical friction is an effective gravitational drag experienced by massive bodies moving through a population of lower mass bodies \citep{Chandrasekhar1949,BinneyTremaine}.  A dark matter subhalo orbiting inside a larger halo will experience dynamical friction due to the density of dark matter particles in the host halo, with a rate
\begin{equation}
\frac{d\bm{v}}{dt} \propto -\frac{G^2 M \rho}{v^3} \bm{v} f(v,\sigma),
\label{DF}
\end{equation}
where $M$ is the mass of the subhalo, $\bm{v}$ is its velocity relative to its host, and $\rho$ is the density through which it moves.  The proportionality constant depends on the distribution function of the particles, as well as the internal structure of the subhalo.  The drag is largest for the most massive subhalos with large $M$.

The effects of dynamical friction are manifest in simulations of nonlinear cosmological structure formation.  Dynamically young halos, like massive galaxy clusters, contain copious amounts of massive substructure that comprise significant fractions of the cluster's total mass.  For example, it is not uncommon to find massive subhalos within clusters, containing $\sim 1-10\%$ of the cluster mass \citep{Wu2013}.  In contrast, massive substructure is rare in dynamically older halos, like those hosting galaxies similar to the Milky Way.  In galactic halos, it is uncommon to find individual subhalos comprising more than $\gtrsim 3\%$ of the host mass \citep{Springel2008}.  Dynamical friction is believed to be the origin of this difference: in the older systems, drag from dynamical friction had sufficient time to cause the orbits of massive subhalos to decay to small radii, where mass loss from tidal stripping and disruption becomes most effective \citep{vandenBosch2016}.  The proportion of substructure in real galaxies and clusters is presumably affected by the same processes that arise in simulations.

Therefore, considerable indirect evidence exists that dynamical friction should operate in actual halos.  It is difficult, however, to directly observe dynamical friction in action, since the relevant timescales are cosmological in duration.  In this paper, we propose a novel method to directly observe the deceleration produced by dynamical friction acting on massive galaxies within galaxy clusters.  The basic idea is that drag from dynamical friction reduces the orbital energy of galaxies within subhalos, which reduces the apocentric radius.  Recent work has shown that a steepening feature in the density profiles of halos occurs near the apocenters of material on its first orbit within halos, termed the splashback feature \citep{Diemer2014,Adhikari2014,More2015}.  In the context of the halo model \citep{Seljak2000,Hu2004}, this feature may be thought of as the boundary of the 1-halo region.  For massive cluster-sized halos, where the 1-halo term is large compared to the 2-halo term, the sharp edge to the 1-halo term produces a steep falloff in the total density profile at the splashback radius $r_{\rm sp}$, and therefore the location of the steepest slope of the density profile occurs close to $r_{\rm sp}$.  Dynamical friction reduces $r_{\rm sp}$, and because friction is more effective for more massive subhalos, high-mass satellites should therefore have a smaller splashback radius than low-mass satellites.

\begin{figure*}
\centerline{\includegraphics[width=\linewidth, trim= 0in 0.5in 0in 0.5in,clip]{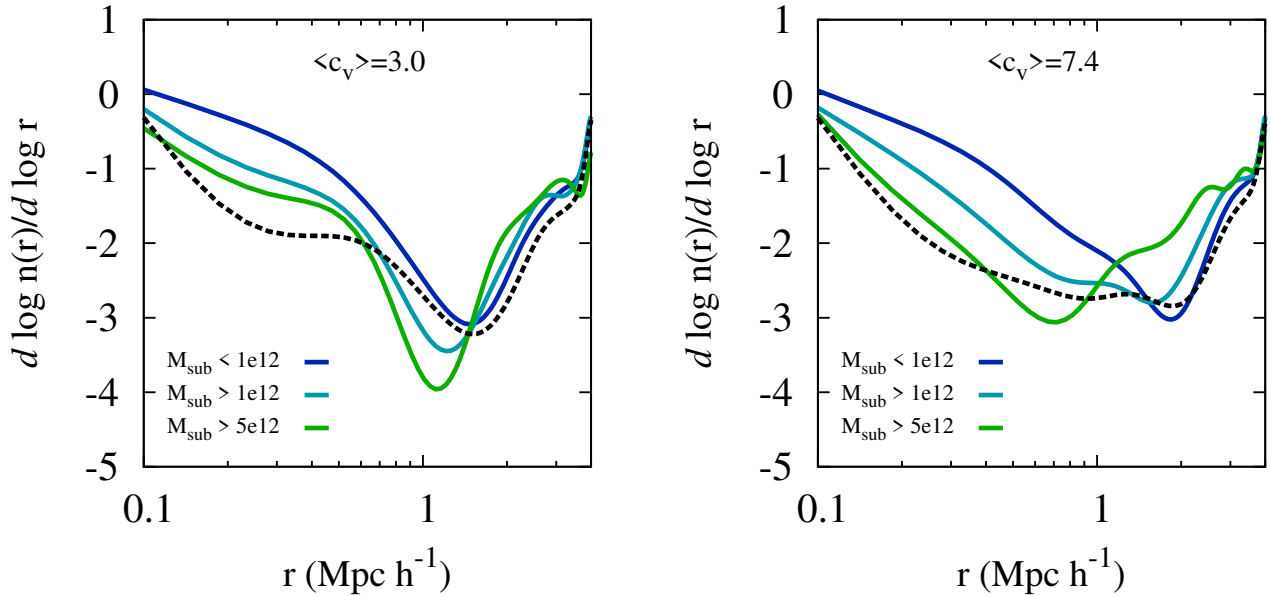}}
\caption{In both panels, the different curves show the logarithmic slope of the number density profile $d\log n/d\log r$ as a function of radius $r$ within cluster-sized halos of virial mass $M_{\rm host}=1-4\times 10^{14} h^{-1} M_\odot$ from the MDPL2 simulation, for various populations within the host halos.  The dashed line corresponds to all dark matter particles, while the solid lines show subhalos of different mass, as denoted in the legend.  The splashback radius occurs at the location of the steepening feature in these profiles.  Subhalos with less than 1\% of the host mass have similar splashback radii as the full set of DM particles, while more massive subhalos have smaller splashback radii, consistent with the effects of dynamical friction.  The left panel shows host halos with $c_{\rm vir}<4$, while the right panel shows host halos with $c_{\rm vir}>6$, illustrating the significant dependence of dynamical friction effects on host concentrations. Subhalo masses are expressed in units of $ h^{-1} M_\odot$.
}
\label{splash}
\end{figure*}

Figure \ref{splash} illustrates this effect, using results from the MDPL2 simulation from the CosmoSim database \citep{Multidark,MultiDark2}.  The figure shows the density slope $d\log n/d\log r$ as a function of radius for subhalos of varying $M_{\rm peak}$, the largest mass attained by each subhalo throughout its history.  Halos and subhalos were selected from the Rockstar \citep{Behroozi13a, Behroozi13b} catalogs publicly available at {\tt http://cosmosim.org}.  The splashback radius for low-mass subhalos is indistinguishable from the splashback radius of dark matter particles, but as $M_{\rm peak}$ increases, the splashback radius steadily decreases.  This effect appears to be caused by dynamical friction, rather than selection effects or other physical processes which affect the spatial distribution of substructure.  
For example, it has long been known that subhalos orbiting at small radii tend to have earlier accretion redshifts than subhalos orbiting at large radii within their hosts \citep{Gao2004,Faltenbacher2006,Contini2012,vandenBosch2016}.  One might therefore imagine that the difference in splashback radii between high-mass and low-mass subhalos might be due to systematic differences in the accretion times for those subhalo samples (caused by resolution effects, for example), but we have checked that the distribution of accretion redshifts as defined in the Rockstar catalog is nearly identical for low-mass and high-mass subhalos.  This precludes the difference in $r_{\rm sp}$ from arising  from mass-dependent selection effects which can convert the radial dependence of mean accretion redshift into an apparent radial dependence of subhalo mass.  Another line of evidence for dynamical friction as the explanation for the trend of $r_{\rm sp}$ with mass is the concentration dependence of the effect.  As Fig.\ \ref{splash} shows, the decrease in $r_{\rm sp}$ is stronger for hosts with higher concentration $c_{\rm vir}$.  This is expected for dynamical friction, since increasing $c_{\rm vir}$ raises the central density, which increases the drag rate as seen in Eqn.\ \eqref{DF}.

The splashback feature therefore offers a direct method to observe the effects of dynamical friction.  It is straightforward to estimate how dynamical friction will affect the splashback radius, using the spherical collapse model of \citet{Adhikari2014}.  We modify their model somewhat, adding an extra term to the equation of motion to account for dynamical friction,
\begin{equation}
\frac{dv_r}{dt} = -\frac{G M(r)}{r^2} - \eta  \frac{G^2 M_{\rm sub} \rho(r)}{|v_r|^3} v_r f(v_r/\sigma).
\label{toy}
\end{equation}
Here, $M_{\rm sub}$ is the mass of the subhalo (we neglect tidal stripping), $v_r$ is its radial velocity, $M(r)$ is the host halo mass enclosed within radius $r$, $\rho(r)$ is the local density at radius $r$, the phase space factor is taken to be that for a Maxwellian distribution, $f(x) = {\rm erf}(x)-2\pi^{-1/2}x\,e^{-x^2/2}$ \citep{BinneyTremaine},  and $\eta$ is the unknown proportionality constant from Eqn.\ \eqref{DF}.  Since we do not have a first principles calculation of $\eta$, we treat it as a free parameter that is fit to the simulation data.  We find that $\eta \approx 1.4$ provides a reasonable fit for the cluster-sized host masses we have considered.  For simplicity, in this toy model we assume radial orbits for subhalos, which is unphysical but reduces the number of dynamical variables.  Because radial orbits pass through the host center $r=0$ where the NFW profile diverges, we instead approximate the host profile using a cored isothermal profile, with $r_{\rm core}=0.1\times r_{\rm vir}$. Figure\ \ref{splash} shows this is a reasonable approximation to the host profile well inside the splashback radius, since at about $0.1 \, r_{\rm sp}$ the density profile slope (black dashed line) rapidly transitions from -2 to 0.

\begin{figure*}
   \centerline{\includegraphics[width=\linewidth, trim= 0in 0.9in 0in 0.5in,clip]{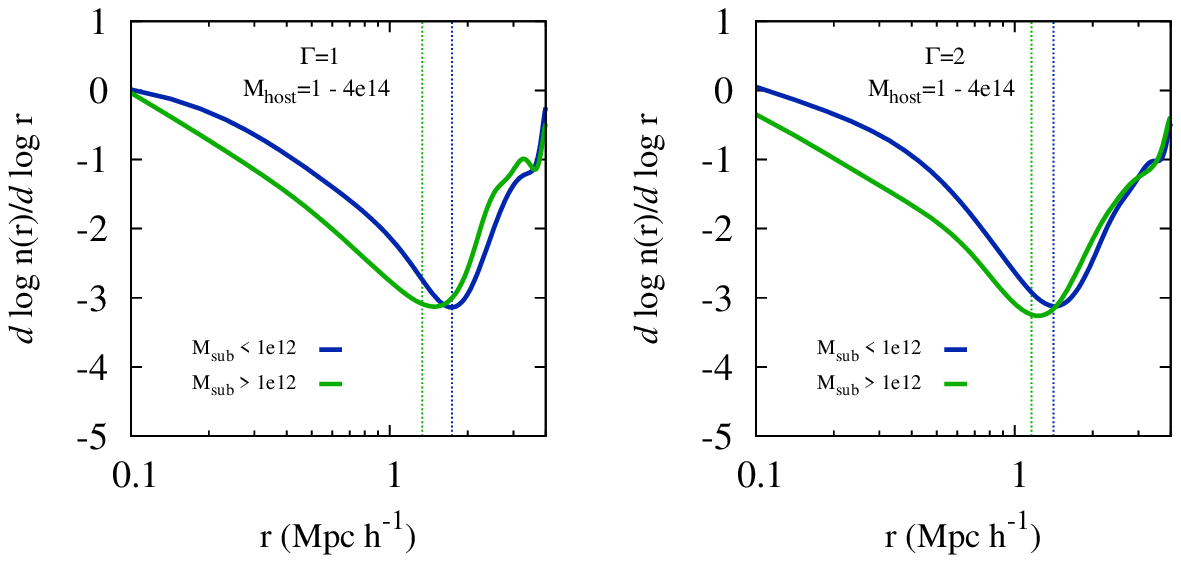}}
   \centerline{\includegraphics[width=\linewidth, trim= 0in 0.5in 0in 0.5in,clip]{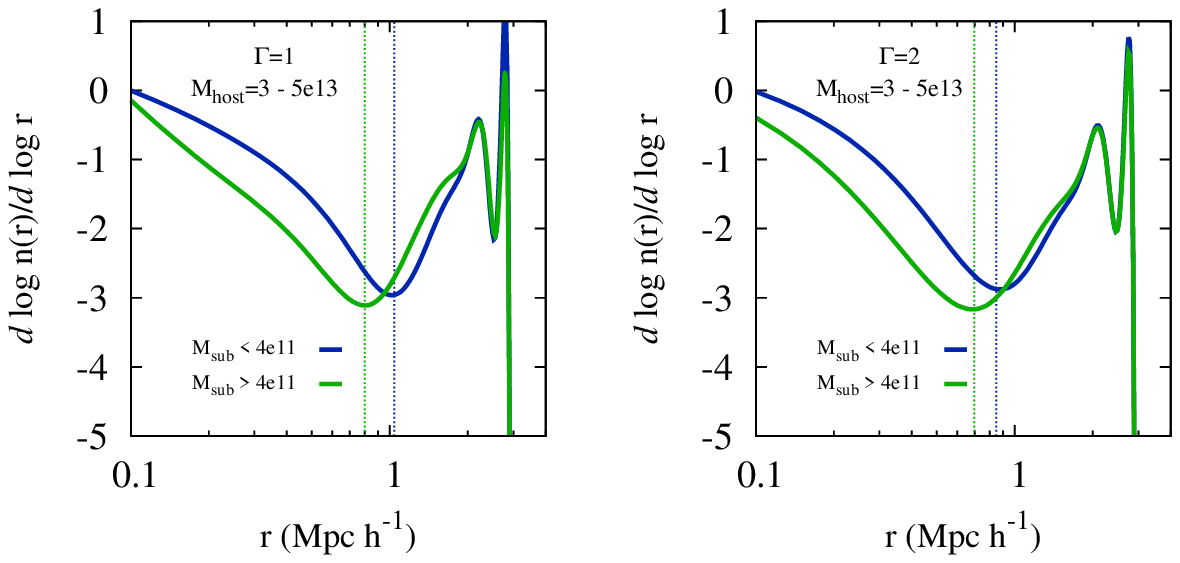}}
\caption{Shift in the location of splashback due to dynamical friction in subhalos with $M_{\rm sub}>0.01M_{\rm host}$ in bins of different accretion rate $\Gamma = d\log M_{\rm host}/d\log a$ (left vs.\ right) and for bins of different host mass $M_{\rm host}$ (top vs.\ bottom). Halo masses are expressed in units of $h^{-1} M_\odot$. The vertical lines show the prediction from the toy model Eqn.\ \eqref{toy}. The blue vertical line shows the predicted location in a model without dynamical friction \citep{Adhikari2014}, which agrees well with the splashback radius $r_{\rm sp}$ for low mass subhalos where dynamical friction is unimportant.  The green vertical line shows the predicted position of splashback from the collapse model with dynamical friction where $\lambda=1.4$, evaluated at the mean subhalo mass of the sample with $M_{\rm sub}>0.01 M_{\rm host}$. 
\label{fig:fit}}
\end{figure*}

This simple toy model reasonably reproduces the location of the splashback feature for different subhalo mass bins and for different host accretion rates $\Gamma = d\log M_{\rm vir}/d\log a$, as shown in Fig.~\ref{fig:fit}. In particular, low-mass subhalos comprising $\lesssim 1\%$ of their host halo's mass do not appear to experience significant drag from dynamical friction.  However, it might be interesting to construct a more realistic model that more accurately describes the structure of the host halo, since the MDPL2 results suggest that dynamical friction effects on the splashback radius depend significantly on the concentration of the host halo (see Fig.\ \ref{splash}).

In principle, therefore,  measurement of the splashback radius for different galaxy masses could be used to constrain the amount of dynamical friction experienced by those galaxies, which in turn constrains the initial masses (at infall) of the subhalos hosting those galaxies.  A comparison of those infall masses to the present-day masses, inferred from galaxy-galaxy lensing of satellites \citep{Sifon2015,Li2016}, then reveals the mass loss from tidal stripping suffered by satellites.  Such a measurement would not only help inform our understanding of dynamics within galaxy clusters, but would also provide new insights on the halo occupation of satellites, directly testing models like subhalo abundance matching \citep{Conroy2006}.  

\begin{figure*}
\centerline{
\includegraphics[width=0.48\textwidth, trim= 0in 0.0in 0in 0.0in,clip]{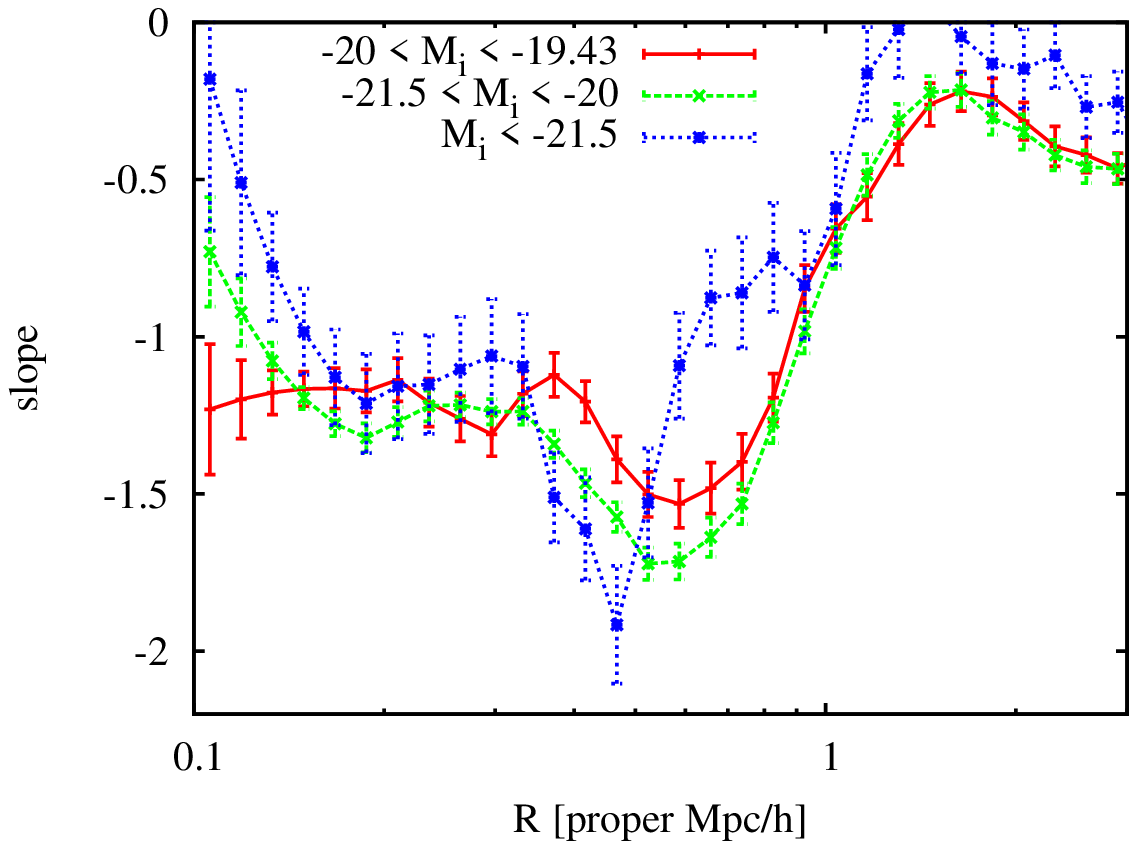}\quad
\includegraphics[width=0.48\textwidth, trim= 0in 0.0in 0in 0.0in,clip]{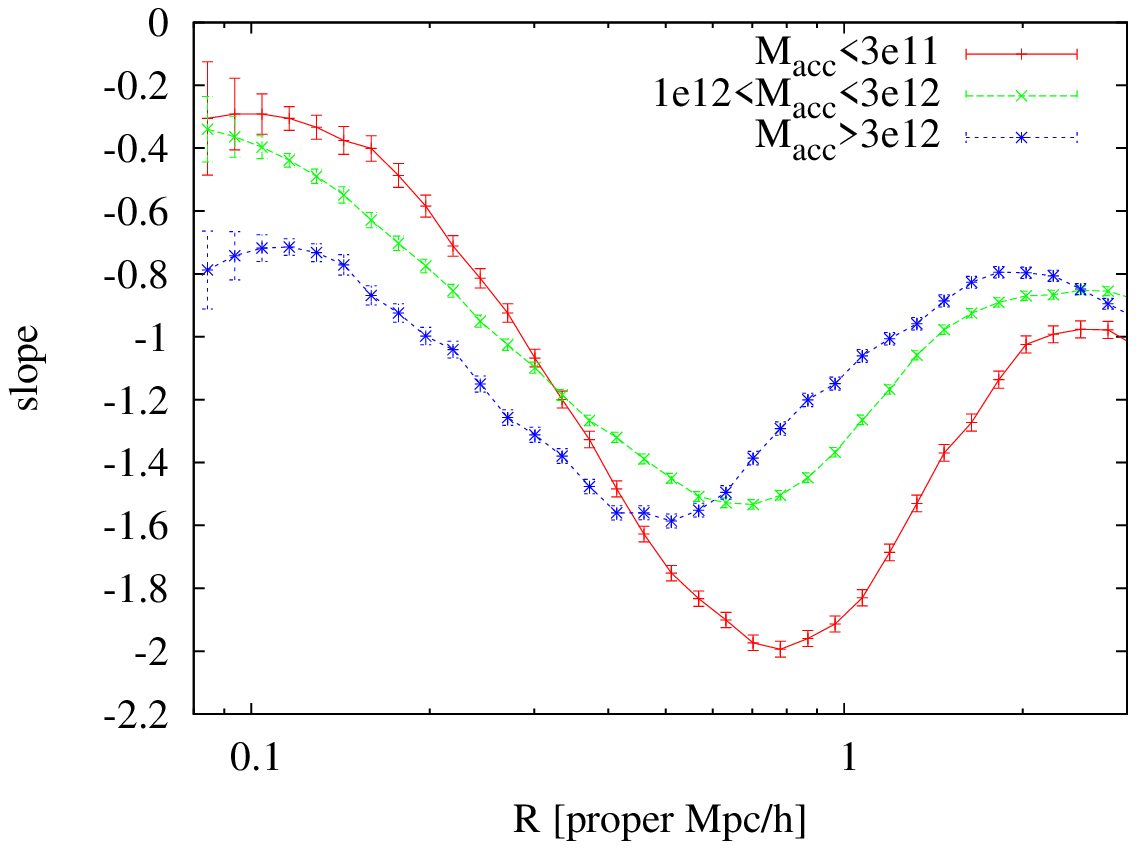}
}
\caption{ (Left) Observations of splashback in clusters from the redMaPPer catalog of galaxy clusters with $0.15<z_c<0.33$ and $10<\lambda<20$. Plotted is the logarithmic slope of the projected number density of galaxies as a function of cluster-centric radius. (Right) Corresponding profiles for subhalos in host halos with $M_{200m}=3-9\times10^{13} h^{-1}M_\odot$ at $z=0.25$ from the MDPL2 simulation.  This host mass range was chosen to match the richness range from the left panel.  Both panels show the slope of the projected density, not the 3D space density, in contrast to the previous figures.
\label{fig:redMaPPer}}
\end{figure*}

\citet{More2016} recently observed the splashback feature in SDSS redMaPPer clusters \citep{Rykoff2014, Rozo2015}.  Building on that work, we conduct a similar measurement to search for signatures of dynamical friction in clusters.  We use redMaPPer clusters with richness $10<\lambda<20$ in the redshift range $0.15<z<0.33$, and compute the projected number density $n(R)$ of SDSS galaxies with measured photometric redshifts.  In Fig.\ \ref{fig:redMaPPer} we plot the logarithmic derivative of the projected number density, $d\log n/d\log R$, for galaxies in various bins of absolute magnitude in the $i$ band, $M_i$.  To determine the absolute magnitude, we ignore the measured photometric redshift for each galaxy, and instead assume the galaxy is at the redshift of the redMaPPer cluster when converting the observed apparent magnitude to absolute magnitude \citep{More2016}.  We also use the angular diameter distance at the cluster redshift to translate angular separation into projected physical separation.  We count galaxies in bins of projected separation $R$, and then compute the logarithmic slope  $d\log n/d\log R$ using a cubic Savitsky-Golay filter over 11 points, accounting for the covariance of the bins.  We have used a lower richness cluster sample than \citet{More2016}, since for a given galaxy population, dynamical friction is expected to be stronger for hosts of lower mass.

As Fig.\ \ref{fig:redMaPPer} shows, the brightest galaxies with $M_i < -21.5$ have a significantly smaller splashback radius than fainter galaxies.  If we assume that $r_{\rm sp}$ is given by the location of the steepest slope of the projected profile (which can somewhat underestimate the value of the 3D splashback radius), we can determine $r_{\rm sp}$ from the Savitsky-Golay fit.  
We find that $r_{\rm sp}=0.42\pm0.04\, h^{-1}\,$Mpc (proper, not comoving) for the $M_i<-21.5$ sample, while $r_{\rm sp}=0.55\pm0.04\, h^{-1}\,$Mpc for the $-19.43<M_i<-20$ sample.   The probability that $r_{\rm sp}$ is smaller for the bright sample than the faint sample is
$\int P(r_{-21.5}) P(r_{-20}) \Theta(r_{-20}-r_{-21.5}) dr_{-21.5} dr_{-20} = 99.3\%$, giving nearly $3\sigma$ confidence that brighter satellites have smaller $r_{\rm sp}$ than fainter satellites in the same clusters.  
This trend qualitatively agrees with our analytical and numerical predictions.  A quantitative comparison with theoretical predictions would require constructing a mock galaxy catalog from simulations and running the redMaPPer algorithm to find clusters and their member galaxies, which is beyond the scope of the present work.  However, we can perform a very crude comparison as follows.  \citet{Simet2016} find that the mean halo mass for richness $\lambda$ is $M_{\rm 200m} = 10^{14.3} h^{-1}M_\odot (\lambda/40)^{1.33}$, while \citet{Farahi2016} find a mean relation $M_{\rm 200c} = 10^{14.19} M_\odot (\lambda/30)^{1.31}$, 
although the scatter about the mean relation appears quite large, $\sim 0.7$ dex at $\lambda \approx 20$.   Assuming this mean relation of \citep{Simet2016}, our sample with $10<\lambda<20$ should roughly correspond to halos with $M_{\rm 200m}=3-9\times 10^{13} h^{-1}M_\odot$.  We have selected halos in this mass range at $z=0.25$ (the median redshift of our sample) from the MDPL2 simulation, and in Fig.\ \ref{fig:redMaPPer} we plot the logarithmic slope of the projected number density, $d\log n/d\log R$, for neighboring halos and subhalos in various bins of $M_{\rm acc}$, the mass at accretion.  (For isolated halos, $M_{\rm acc}=M_{\rm vir}$.)  As the figure shows, subhalos with $M_{\rm acc}\gtrsim 3\times 10^{12} h^{-1}M_\odot$ have splashback radii comparable to our bright galaxy sample with $M_i<-21.5$, significantly smaller than $r_{\rm sp}$ for subhalos with lower $M_{\rm acc}$.   Therefore, we would expect bright redMaPPer galaxies to have infall masses of this order, and a comparison with their present-day lensing masses \citep{Sifon2015,Li2016} would be an interesting test of whether dynamical friction in real clusters proceeds similarly to the behavior found in simulations.

Besides probing the subhalo occupation distribution of cluster satellites, 
observations of dynamical friction can also help constrain more fundamental physics.  We have estimated the amount of dynamical friction that occurs within cold dark matter (CDM) cosmologies, in which gravity follows Einstein's general relativity and dark matter consists of collisionless particles.  Departures from the CDM model could produce significant deviations in the splashback radius of massive galaxies.  For example, some previous works have found that dynamical friction can be significantly strengthened in MOND-like theories without particle dark matter \citep{Nipoti2008}, which may be interesting to explore in more recent variants of modified gravity models \citep{Berezhiani2015}.  Similarly, certain self-interacting dark matter models can also predict an effective drag of satellites \citep{Kahlhoefer2014} that could be studied using the splashback feature.  Given the detection of the splashback feature in SDSS clusters, and improvements from deeper imaging surveys like DES and HSC, splashback may provide a new window onto a variety of physics.

\begin{acknowledgments}
We thank Bhuvnesh Jain, Surhud More and Scott Tremaine for helpful discussions.
This work was supported by NASA under grant HST-AR-14291.001-A from the Space Telescope Science Institute, which is operated by the Association of Universities for Research in Astronomy, Inc., under NASA contract NAS 5-26555.  ND was also supported by the Institute for Advanced Study and the Ambrose Monell Foundation.
The CosmoSim database used in this paper is a service by the Leibniz-Institute for Astrophysics Potsdam (AIP).
The MultiDark database was developed in cooperation with the Spanish MultiDark Consolider Project CSD2009-00064.
The MultiDark-Planck (MDPL) and the BigMD simulation suite have been performed in the Supermuc supercomputer at LRZ using time granted by PRACE.
\end{acknowledgments}

\newcommand{\mnras}{MNRAS}
\newcommand{\physrep}{Physics Reports}

\bibliography{friction}

\begin{thebibliography}{28}%
\makeatletter
\providecommand \@ifxundefined [1]{%
 \@ifx{#1\undefined}
}%
\providecommand \@ifnum [1]{%
 \ifnum #1\expandafter \@firstoftwo
 \else \expandafter \@secondoftwo
 \fi
}%
\providecommand \@ifx [1]{%
 \ifx #1\expandafter \@firstoftwo
 \else \expandafter \@secondoftwo
 \fi
}%
\providecommand \natexlab [1]{#1}%
\providecommand \enquote  [1]{``#1''}%
\providecommand \bibnamefont  [1]{#1}%
\providecommand \bibfnamefont [1]{#1}%
\providecommand \citenamefont [1]{#1}%
\providecommand \href@noop [0]{\@secondoftwo}%
\providecommand \href [0]{\begingroup \@sanitize@url \@href}%
\providecommand \@href[1]{\@@startlink{#1}\@@href}%
\providecommand \@@href[1]{\endgroup#1\@@endlink}%
\providecommand \@sanitize@url [0]{\catcode `\\12\catcode `\$12\catcode
  `\&12\catcode `\#12\catcode `\^12\catcode `\_12\catcode `\%12\relax}%
\providecommand \@@startlink[1]{}%
\providecommand \@@endlink[0]{}%
\providecommand \url  [0]{\begingroup\@sanitize@url \@url }%
\providecommand \@url [1]{\endgroup\@href {#1}{\urlprefix }}%
\providecommand \urlprefix  [0]{URL }%
\providecommand \Eprint [0]{\href }%
\providecommand \doibase [0]{http://dx.doi.org/}%
\providecommand \selectlanguage [0]{\@gobble}%
\providecommand \bibinfo  [0]{\@secondoftwo}%
\providecommand \bibfield  [0]{\@secondoftwo}%
\providecommand \translation [1]{[#1]}%
\providecommand \BibitemOpen [0]{}%
\providecommand \bibitemStop [0]{}%
\providecommand \bibitemNoStop [0]{.\EOS\space}%
\providecommand \EOS [0]{\spacefactor3000\relax}%
\providecommand \BibitemShut  [1]{\csname bibitem#1\endcsname}%
\let\auto@bib@innerbib\@empty
\bibitem [{\citenamefont {{Chandrasekhar}}(1949)}]{Chandrasekhar1949}%
  \BibitemOpen
  \bibfield  {author} {\bibinfo {author} {\bibfnamefont {S.}~\bibnamefont
  {{Chandrasekhar}}},\ }\href {\doibase 10.1103/RevModPhys.21.383} {\bibfield
  {journal} {\bibinfo  {journal} {Reviews of Modern Physics}\ }\textbf
  {\bibinfo {volume} {21}},\ \bibinfo {pages} {383} (\bibinfo {year}
  {1949})}\BibitemShut {NoStop}%
\bibitem [{\citenamefont {{Binney}}\ and\ \citenamefont
  {{Tremaine}}(2008)}]{BinneyTremaine}%
  \BibitemOpen
  \bibfield  {author} {\bibinfo {author} {\bibfnamefont {J.}~\bibnamefont
  {{Binney}}}\ and\ \bibinfo {author} {\bibfnamefont {S.}~\bibnamefont
  {{Tremaine}}},\ }\href@noop {} {\emph {\bibinfo {title} {Galactic Dynamics:
  Second Edition, by James Binney and Scott Tremaine.~ISBN 978-0-691-13026-2
  (HB).~Published by Princeton University Press, Princeton, NJ USA, 2008.}}},\
  edited by\ \bibinfo {editor} {\bibfnamefont {J.}~\bibnamefont {{Binney}}}\
  and\ \bibinfo {editor} {\bibfnamefont {S.}~\bibnamefont {{Tremaine}}}\
  (\bibinfo  {publisher} {Princeton University Press},\ \bibinfo {year}
  {2008})\BibitemShut {NoStop}%
\bibitem [{\citenamefont {{Wu}}\ \emph {et~al.}(2013)\citenamefont {{Wu}},
  \citenamefont {{Hahn}}, \citenamefont {{Wechsler}}, \citenamefont
  {{Behroozi}},\ and\ \citenamefont {{Mao}}}]{Wu2013}%
  \BibitemOpen
  \bibfield  {author} {\bibinfo {author} {\bibfnamefont {H.-Y.}\ \bibnamefont
  {{Wu}}}, \bibinfo {author} {\bibfnamefont {O.}~\bibnamefont {{Hahn}}},
  \bibinfo {author} {\bibfnamefont {R.~H.}\ \bibnamefont {{Wechsler}}},
  \bibinfo {author} {\bibfnamefont {P.~S.}\ \bibnamefont {{Behroozi}}}, \ and\
  \bibinfo {author} {\bibfnamefont {Y.-Y.}\ \bibnamefont {{Mao}}},\ }\href
  {\doibase 10.1088/0004-637X/767/1/23} {\bibfield  {journal} {\bibinfo
  {journal} {\apj}\ }\textbf {\bibinfo {volume} {767}},\ \bibinfo {eid} {23}
  (\bibinfo {year} {2013})},\ \Eprint {http://arxiv.org/abs/1210.6358}
  {arXiv:1210.6358 [astro-ph.CO]} \BibitemShut {NoStop}%
\bibitem [{\citenamefont {{Springel}}\ \emph {et~al.}(2008)\citenamefont
  {{Springel}}, \citenamefont {{Wang}}, \citenamefont {{Vogelsberger}},
  \citenamefont {{Ludlow}}, \citenamefont {{Jenkins}}, \citenamefont {{Helmi}},
  \citenamefont {{Navarro}}, \citenamefont {{Frenk}},\ and\ \citenamefont
  {{White}}}]{Springel2008}%
  \BibitemOpen
  \bibfield  {author} {\bibinfo {author} {\bibfnamefont {V.}~\bibnamefont
  {{Springel}}}, \bibinfo {author} {\bibfnamefont {J.}~\bibnamefont {{Wang}}},
  \bibinfo {author} {\bibfnamefont {M.}~\bibnamefont {{Vogelsberger}}},
  \bibinfo {author} {\bibfnamefont {A.}~\bibnamefont {{Ludlow}}}, \bibinfo
  {author} {\bibfnamefont {A.}~\bibnamefont {{Jenkins}}}, \bibinfo {author}
  {\bibfnamefont {A.}~\bibnamefont {{Helmi}}}, \bibinfo {author} {\bibfnamefont
  {J.~F.}\ \bibnamefont {{Navarro}}}, \bibinfo {author} {\bibfnamefont {C.~S.}\
  \bibnamefont {{Frenk}}}, \ and\ \bibinfo {author} {\bibfnamefont {S.~D.~M.}\
  \bibnamefont {{White}}},\ }\href {\doibase 10.1111/j.1365-2966.2008.14066.x}
  {\bibfield  {journal} {\bibinfo  {journal} {\mnras}\ }\textbf {\bibinfo
  {volume} {391}},\ \bibinfo {pages} {1685} (\bibinfo {year} {2008})},\ \Eprint
  {http://arxiv.org/abs/0809.0898} {arXiv:0809.0898} \BibitemShut {NoStop}%
\bibitem [{\citenamefont {{van den Bosch}}\ \emph {et~al.}(2016)\citenamefont
  {{van den Bosch}}, \citenamefont {{Jiang}}, \citenamefont {{Campbell}},\ and\
  \citenamefont {{Behroozi}}}]{vandenBosch2016}%
  \BibitemOpen
  \bibfield  {author} {\bibinfo {author} {\bibfnamefont {F.~C.}\ \bibnamefont
  {{van den Bosch}}}, \bibinfo {author} {\bibfnamefont {F.}~\bibnamefont
  {{Jiang}}}, \bibinfo {author} {\bibfnamefont {D.}~\bibnamefont {{Campbell}}},
  \ and\ \bibinfo {author} {\bibfnamefont {P.}~\bibnamefont {{Behroozi}}},\
  }\href {\doibase 10.1093/mnras/stv2338} {\bibfield  {journal} {\bibinfo
  {journal} {\mnras}\ }\textbf {\bibinfo {volume} {455}},\ \bibinfo {pages}
  {158} (\bibinfo {year} {2016})},\ \Eprint {http://arxiv.org/abs/1510.01586}
  {arXiv:1510.01586} \BibitemShut {NoStop}%
\bibitem [{\citenamefont {{Diemer}}\ and\ \citenamefont
  {{Kravtsov}}(2014)}]{Diemer2014}%
  \BibitemOpen
  \bibfield  {author} {\bibinfo {author} {\bibfnamefont {B.}~\bibnamefont
  {{Diemer}}}\ and\ \bibinfo {author} {\bibfnamefont {A.~V.}\ \bibnamefont
  {{Kravtsov}}},\ }\href {\doibase 10.1088/0004-637X/789/1/1} {\bibfield
  {journal} {\bibinfo  {journal} {\apj}\ }\textbf {\bibinfo {volume} {789}},\
  \bibinfo {eid} {1} (\bibinfo {year} {2014})},\ \Eprint
  {http://arxiv.org/abs/1401.1216} {arXiv:1401.1216} \BibitemShut {NoStop}%
\bibitem [{\citenamefont {{Adhikari}}\ \emph {et~al.}(2014)\citenamefont
  {{Adhikari}}, \citenamefont {{Dalal}},\ and\ \citenamefont
  {{Chamberlain}}}]{Adhikari2014}%
  \BibitemOpen
  \bibfield  {author} {\bibinfo {author} {\bibfnamefont {S.}~\bibnamefont
  {{Adhikari}}}, \bibinfo {author} {\bibfnamefont {N.}~\bibnamefont {{Dalal}}},
  \ and\ \bibinfo {author} {\bibfnamefont {R.~T.}\ \bibnamefont
  {{Chamberlain}}},\ }\href {\doibase 10.1088/1475-7516/2014/11/019} {\bibfield
   {journal} {\bibinfo  {journal} {JCAP}\ }\textbf {\bibinfo {volume} {11}},\
  \bibinfo {eid} {019} (\bibinfo {year} {2014})},\ \Eprint
  {http://arxiv.org/abs/1409.4482} {arXiv:1409.4482} \BibitemShut {NoStop}%
\bibitem [{\citenamefont {{More}}\ \emph {et~al.}(2015)\citenamefont {{More}},
  \citenamefont {{Diemer}},\ and\ \citenamefont {{Kravtsov}}}]{More2015}%
  \BibitemOpen
  \bibfield  {author} {\bibinfo {author} {\bibfnamefont {S.}~\bibnamefont
  {{More}}}, \bibinfo {author} {\bibfnamefont {B.}~\bibnamefont {{Diemer}}}, \
  and\ \bibinfo {author} {\bibfnamefont {A.~V.}\ \bibnamefont {{Kravtsov}}},\
  }\href {\doibase 10.1088/0004-637X/810/1/36} {\bibfield  {journal} {\bibinfo
  {journal} {\apj}\ }\textbf {\bibinfo {volume} {810}},\ \bibinfo {eid} {36}
  (\bibinfo {year} {2015})},\ \Eprint {http://arxiv.org/abs/1504.05591}
  {arXiv:1504.05591} \BibitemShut {NoStop}%
\bibitem [{\citenamefont {{Seljak}}(2000)}]{Seljak2000}%
  \BibitemOpen
  \bibfield  {author} {\bibinfo {author} {\bibfnamefont {U.}~\bibnamefont
  {{Seljak}}},\ }\href {\doibase 10.1046/j.1365-8711.2000.03715.x} {\bibfield
  {journal} {\bibinfo  {journal} {\mnras}\ }\textbf {\bibinfo {volume} {318}},\
  \bibinfo {pages} {203} (\bibinfo {year} {2000})},\ \Eprint
  {http://arxiv.org/abs/astro-ph/0001493} {astro-ph/0001493} \BibitemShut
  {NoStop}%
\bibitem [{\citenamefont {{Hu}}\ and\ \citenamefont {{Jain}}(2004)}]{Hu2004}%
  \BibitemOpen
  \bibfield  {author} {\bibinfo {author} {\bibfnamefont {W.}~\bibnamefont
  {{Hu}}}\ and\ \bibinfo {author} {\bibfnamefont {B.}~\bibnamefont {{Jain}}},\
  }\href {\doibase 10.1103/PhysRevD.70.043009} {\bibfield  {journal} {\bibinfo
  {journal} {\prd}\ }\textbf {\bibinfo {volume} {70}},\ \bibinfo {eid} {043009}
  (\bibinfo {year} {2004})},\ \Eprint {http://arxiv.org/abs/astro-ph/0312395}
  {astro-ph/0312395} \BibitemShut {NoStop}%
\bibitem [{\citenamefont {{Riebe}}\ \emph {et~al.}(2013)\citenamefont
  {{Riebe}}, \citenamefont {{Partl}}, \citenamefont {{Enke}}, \citenamefont
  {{Forero-Romero}}, \citenamefont {{Gottl{\"o}ber}}, \citenamefont {{Klypin}},
  \citenamefont {{Lemson}}, \citenamefont {{Prada}}, \citenamefont {{Primack}},
  \citenamefont {{Steinmetz}},\ and\ \citenamefont
  {{Turchaninov}}}]{Multidark}%
  \BibitemOpen
  \bibfield  {author} {\bibinfo {author} {\bibfnamefont {K.}~\bibnamefont
  {{Riebe}}}, \bibinfo {author} {\bibfnamefont {A.~M.}\ \bibnamefont
  {{Partl}}}, \bibinfo {author} {\bibfnamefont {H.}~\bibnamefont {{Enke}}},
  \bibinfo {author} {\bibfnamefont {J.}~\bibnamefont {{Forero-Romero}}},
  \bibinfo {author} {\bibfnamefont {S.}~\bibnamefont {{Gottl{\"o}ber}}},
  \bibinfo {author} {\bibfnamefont {A.}~\bibnamefont {{Klypin}}}, \bibinfo
  {author} {\bibfnamefont {G.}~\bibnamefont {{Lemson}}}, \bibinfo {author}
  {\bibfnamefont {F.}~\bibnamefont {{Prada}}}, \bibinfo {author} {\bibfnamefont
  {J.~R.}\ \bibnamefont {{Primack}}}, \bibinfo {author} {\bibfnamefont
  {M.}~\bibnamefont {{Steinmetz}}}, \ and\ \bibinfo {author} {\bibfnamefont
  {V.}~\bibnamefont {{Turchaninov}}},\ }\href {\doibase 10.1002/asna.201211900}
  {\bibfield  {journal} {\bibinfo  {journal} {Astronomische Nachrichten}\
  }\textbf {\bibinfo {volume} {334}},\ \bibinfo {pages} {691} (\bibinfo {year}
  {2013})}\BibitemShut {NoStop}%
\bibitem [{\citenamefont {{Klypin}}\ \emph {et~al.}(2016)\citenamefont
  {{Klypin}}, \citenamefont {{Yepes}}, \citenamefont {{Gottl{\"o}ber}},
  \citenamefont {{Prada}},\ and\ \citenamefont {{He{\ss}}}}]{MultiDark2}%
  \BibitemOpen
  \bibfield  {author} {\bibinfo {author} {\bibfnamefont {A.}~\bibnamefont
  {{Klypin}}}, \bibinfo {author} {\bibfnamefont {G.}~\bibnamefont {{Yepes}}},
  \bibinfo {author} {\bibfnamefont {S.}~\bibnamefont {{Gottl{\"o}ber}}},
  \bibinfo {author} {\bibfnamefont {F.}~\bibnamefont {{Prada}}}, \ and\
  \bibinfo {author} {\bibfnamefont {S.}~\bibnamefont {{He{\ss}}}},\ }\href
  {\doibase 10.1093/mnras/stw248} {\bibfield  {journal} {\bibinfo  {journal}
  {\mnras}\ }\textbf {\bibinfo {volume} {457}},\ \bibinfo {pages} {4340}
  (\bibinfo {year} {2016})}\BibitemShut {NoStop}%
\bibitem [{\citenamefont {{Behroozi}}\ \emph
  {et~al.}(2013{\natexlab{a}})\citenamefont {{Behroozi}}, \citenamefont
  {{Wechsler}},\ and\ \citenamefont {{Wu}}}]{Behroozi13a}%
  \BibitemOpen
  \bibfield  {author} {\bibinfo {author} {\bibfnamefont {P.~S.}\ \bibnamefont
  {{Behroozi}}}, \bibinfo {author} {\bibfnamefont {R.~H.}\ \bibnamefont
  {{Wechsler}}}, \ and\ \bibinfo {author} {\bibfnamefont {H.-Y.}\ \bibnamefont
  {{Wu}}},\ }\href {\doibase 10.1088/0004-637X/762/2/109} {\bibfield  {journal}
  {\bibinfo  {journal} {\apj}\ }\textbf {\bibinfo {volume} {762}},\ \bibinfo
  {eid} {109} (\bibinfo {year} {2013}{\natexlab{a}})},\ \Eprint
  {http://arxiv.org/abs/1110.4372} {arXiv:1110.4372 [astro-ph.CO]} \BibitemShut
  {NoStop}%
\bibitem [{\citenamefont {{Behroozi}}\ \emph
  {et~al.}(2013{\natexlab{b}})\citenamefont {{Behroozi}}, \citenamefont
  {{Wechsler}}, \citenamefont {{Wu}}, \citenamefont {{Busha}}, \citenamefont
  {{Klypin}},\ and\ \citenamefont {{Primack}}}]{Behroozi13b}%
  \BibitemOpen
  \bibfield  {author} {\bibinfo {author} {\bibfnamefont {P.~S.}\ \bibnamefont
  {{Behroozi}}}, \bibinfo {author} {\bibfnamefont {R.~H.}\ \bibnamefont
  {{Wechsler}}}, \bibinfo {author} {\bibfnamefont {H.-Y.}\ \bibnamefont
  {{Wu}}}, \bibinfo {author} {\bibfnamefont {M.~T.}\ \bibnamefont {{Busha}}},
  \bibinfo {author} {\bibfnamefont {A.~A.}\ \bibnamefont {{Klypin}}}, \ and\
  \bibinfo {author} {\bibfnamefont {J.~R.}\ \bibnamefont {{Primack}}},\ }\href
  {\doibase 10.1088/0004-637X/763/1/18} {\bibfield  {journal} {\bibinfo
  {journal} {\apj}\ }\textbf {\bibinfo {volume} {763}},\ \bibinfo {eid} {18}
  (\bibinfo {year} {2013}{\natexlab{b}})},\ \Eprint
  {http://arxiv.org/abs/1110.4370} {arXiv:1110.4370 [astro-ph.CO]} \BibitemShut
  {NoStop}%
\bibitem [{\citenamefont {{Gao}}\ \emph {et~al.}(2004)\citenamefont {{Gao}},
  \citenamefont {{White}}, \citenamefont {{Jenkins}}, \citenamefont
  {{Stoehr}},\ and\ \citenamefont {{Springel}}}]{Gao2004}%
  \BibitemOpen
  \bibfield  {author} {\bibinfo {author} {\bibfnamefont {L.}~\bibnamefont
  {{Gao}}}, \bibinfo {author} {\bibfnamefont {S.~D.~M.}\ \bibnamefont
  {{White}}}, \bibinfo {author} {\bibfnamefont {A.}~\bibnamefont {{Jenkins}}},
  \bibinfo {author} {\bibfnamefont {F.}~\bibnamefont {{Stoehr}}}, \ and\
  \bibinfo {author} {\bibfnamefont {V.}~\bibnamefont {{Springel}}},\ }\href
  {\doibase 10.1111/j.1365-2966.2004.08360.x} {\bibfield  {journal} {\bibinfo
  {journal} {\mnras}\ }\textbf {\bibinfo {volume} {355}},\ \bibinfo {pages}
  {819} (\bibinfo {year} {2004})},\ \Eprint
  {http://arxiv.org/abs/astro-ph/0404589} {astro-ph/0404589} \BibitemShut
  {NoStop}%
\bibitem [{\citenamefont {{Faltenbacher}}\ and\ \citenamefont
  {{Diemand}}(2006)}]{Faltenbacher2006}%
  \BibitemOpen
  \bibfield  {author} {\bibinfo {author} {\bibfnamefont {A.}~\bibnamefont
  {{Faltenbacher}}}\ and\ \bibinfo {author} {\bibfnamefont {J.}~\bibnamefont
  {{Diemand}}},\ }\href {\doibase 10.1111/j.1365-2966.2006.10421.x} {\bibfield
  {journal} {\bibinfo  {journal} {\mnras}\ }\textbf {\bibinfo {volume} {369}},\
  \bibinfo {pages} {1698} (\bibinfo {year} {2006})},\ \Eprint
  {http://arxiv.org/abs/astro-ph/0602197} {astro-ph/0602197} \BibitemShut
  {NoStop}%
\bibitem [{\citenamefont {{Contini}}\ \emph {et~al.}(2012)\citenamefont
  {{Contini}}, \citenamefont {{De Lucia}},\ and\ \citenamefont
  {{Borgani}}}]{Contini2012}%
  \BibitemOpen
  \bibfield  {author} {\bibinfo {author} {\bibfnamefont {E.}~\bibnamefont
  {{Contini}}}, \bibinfo {author} {\bibfnamefont {G.}~\bibnamefont {{De
  Lucia}}}, \ and\ \bibinfo {author} {\bibfnamefont {S.}~\bibnamefont
  {{Borgani}}},\ }\href {\doibase 10.1111/j.1365-2966.2011.20149.x} {\bibfield
  {journal} {\bibinfo  {journal} {\mnras}\ }\textbf {\bibinfo {volume} {420}},\
  \bibinfo {pages} {2978} (\bibinfo {year} {2012})},\ \Eprint
  {http://arxiv.org/abs/1111.1911} {arXiv:1111.1911} \BibitemShut {NoStop}%
\bibitem [{\citenamefont {{Sif{\'o}n}}\ \emph {et~al.}(2015)\citenamefont
  {{Sif{\'o}n}}, \citenamefont {{Cacciato}}, \citenamefont {{Hoekstra}},
  \citenamefont {{Brouwer}}, \citenamefont {{van Uitert}}, \citenamefont
  {{Viola}}, \citenamefont {{Baldry}}, \citenamefont {{Brough}}, \citenamefont
  {{Brown}}, \citenamefont {{Choi}}, \citenamefont {{Driver}}, \citenamefont
  {{Erben}}, \citenamefont {{Grado}}, \citenamefont {{Heymans}}, \citenamefont
  {{Hildebrandt}}, \citenamefont {{Joachimi}}, \citenamefont {{de Jong}},
  \citenamefont {{Kuijken}}, \citenamefont {{McFarland}}, \citenamefont
  {{Miller}}, \citenamefont {{Nakajima}}, \citenamefont {{Napolitano}},
  \citenamefont {{Norberg}}, \citenamefont {{Robotham}}, \citenamefont
  {{Schneider}},\ and\ \citenamefont {{Kleijn}}}]{Sifon2015}%
  \BibitemOpen
  \bibfield  {author} {\bibinfo {author} {\bibfnamefont {C.}~\bibnamefont
  {{Sif{\'o}n}}}, \bibinfo {author} {\bibfnamefont {M.}~\bibnamefont
  {{Cacciato}}}, \bibinfo {author} {\bibfnamefont {H.}~\bibnamefont
  {{Hoekstra}}}, \bibinfo {author} {\bibfnamefont {M.}~\bibnamefont
  {{Brouwer}}}, \bibinfo {author} {\bibfnamefont {E.}~\bibnamefont {{van
  Uitert}}}, \bibinfo {author} {\bibfnamefont {M.}~\bibnamefont {{Viola}}},
  \bibinfo {author} {\bibfnamefont {I.}~\bibnamefont {{Baldry}}}, \bibinfo
  {author} {\bibfnamefont {S.}~\bibnamefont {{Brough}}}, \bibinfo {author}
  {\bibfnamefont {M.~J.~I.}\ \bibnamefont {{Brown}}}, \bibinfo {author}
  {\bibfnamefont {A.}~\bibnamefont {{Choi}}}, \bibinfo {author} {\bibfnamefont
  {S.~P.}\ \bibnamefont {{Driver}}}, \bibinfo {author} {\bibfnamefont
  {T.}~\bibnamefont {{Erben}}}, \bibinfo {author} {\bibfnamefont
  {A.}~\bibnamefont {{Grado}}}, \bibinfo {author} {\bibfnamefont
  {C.}~\bibnamefont {{Heymans}}}, \bibinfo {author} {\bibfnamefont
  {H.}~\bibnamefont {{Hildebrandt}}}, \bibinfo {author} {\bibfnamefont
  {B.}~\bibnamefont {{Joachimi}}}, \bibinfo {author} {\bibfnamefont {J.~T.~A.}\
  \bibnamefont {{de Jong}}}, \bibinfo {author} {\bibfnamefont {K.}~\bibnamefont
  {{Kuijken}}}, \bibinfo {author} {\bibfnamefont {J.}~\bibnamefont
  {{McFarland}}}, \bibinfo {author} {\bibfnamefont {L.}~\bibnamefont
  {{Miller}}}, \bibinfo {author} {\bibfnamefont {R.}~\bibnamefont
  {{Nakajima}}}, \bibinfo {author} {\bibfnamefont {N.}~\bibnamefont
  {{Napolitano}}}, \bibinfo {author} {\bibfnamefont {P.}~\bibnamefont
  {{Norberg}}}, \bibinfo {author} {\bibfnamefont {A.~S.~G.}\ \bibnamefont
  {{Robotham}}}, \bibinfo {author} {\bibfnamefont {P.}~\bibnamefont
  {{Schneider}}}, \ and\ \bibinfo {author} {\bibfnamefont {G.~V.}\ \bibnamefont
  {{Kleijn}}},\ }\href {\doibase 10.1093/mnras/stv2051} {\bibfield  {journal}
  {\bibinfo  {journal} {\mnras}\ }\textbf {\bibinfo {volume} {454}},\ \bibinfo
  {pages} {3938} (\bibinfo {year} {2015})},\ \Eprint
  {http://arxiv.org/abs/1507.00737} {arXiv:1507.00737} \BibitemShut {NoStop}%
\bibitem [{\citenamefont {{Li}}\ \emph {et~al.}(2016)\citenamefont {{Li}},
  \citenamefont {{Shan}}, \citenamefont {{Kneib}}, \citenamefont {{Mo}},
  \citenamefont {{Rozo}}, \citenamefont {{Leauthaud}}, \citenamefont
  {{Moustakas}}, \citenamefont {{Xie}}, \citenamefont {{Erben}}, \citenamefont
  {{Van Waerbeke}}, \citenamefont {{Makler}}, \citenamefont {{Rykoff}},\ and\
  \citenamefont {{Moraes}}}]{Li2016}%
  \BibitemOpen
  \bibfield  {author} {\bibinfo {author} {\bibfnamefont {R.}~\bibnamefont
  {{Li}}}, \bibinfo {author} {\bibfnamefont {H.}~\bibnamefont {{Shan}}},
  \bibinfo {author} {\bibfnamefont {J.-P.}\ \bibnamefont {{Kneib}}}, \bibinfo
  {author} {\bibfnamefont {H.}~\bibnamefont {{Mo}}}, \bibinfo {author}
  {\bibfnamefont {E.}~\bibnamefont {{Rozo}}}, \bibinfo {author} {\bibfnamefont
  {A.}~\bibnamefont {{Leauthaud}}}, \bibinfo {author} {\bibfnamefont
  {J.}~\bibnamefont {{Moustakas}}}, \bibinfo {author} {\bibfnamefont
  {L.}~\bibnamefont {{Xie}}}, \bibinfo {author} {\bibfnamefont
  {T.}~\bibnamefont {{Erben}}}, \bibinfo {author} {\bibfnamefont
  {L.}~\bibnamefont {{Van Waerbeke}}}, \bibinfo {author} {\bibfnamefont
  {M.}~\bibnamefont {{Makler}}}, \bibinfo {author} {\bibfnamefont
  {E.}~\bibnamefont {{Rykoff}}}, \ and\ \bibinfo {author} {\bibfnamefont
  {B.}~\bibnamefont {{Moraes}}},\ }\href {\doibase 10.1093/mnras/stw494}
  {\bibfield  {journal} {\bibinfo  {journal} {\mnras}\ }\textbf {\bibinfo
  {volume} {458}},\ \bibinfo {pages} {2573} (\bibinfo {year} {2016})},\ \Eprint
  {http://arxiv.org/abs/1507.01464} {arXiv:1507.01464} \BibitemShut {NoStop}%
\bibitem [{\citenamefont {{Conroy}}\ \emph {et~al.}(2006)\citenamefont
  {{Conroy}}, \citenamefont {{Wechsler}},\ and\ \citenamefont
  {{Kravtsov}}}]{Conroy2006}%
  \BibitemOpen
  \bibfield  {author} {\bibinfo {author} {\bibfnamefont {C.}~\bibnamefont
  {{Conroy}}}, \bibinfo {author} {\bibfnamefont {R.~H.}\ \bibnamefont
  {{Wechsler}}}, \ and\ \bibinfo {author} {\bibfnamefont {A.~V.}\ \bibnamefont
  {{Kravtsov}}},\ }\href {\doibase 10.1086/503602} {\bibfield  {journal}
  {\bibinfo  {journal} {\apj}\ }\textbf {\bibinfo {volume} {647}},\ \bibinfo
  {pages} {201} (\bibinfo {year} {2006})},\ \Eprint
  {http://arxiv.org/abs/astro-ph/0512234} {astro-ph/0512234} \BibitemShut
  {NoStop}%
\bibitem [{\citenamefont {{More}}\ \emph {et~al.}(2016)\citenamefont {{More}},
  \citenamefont {{Miyatake}}, \citenamefont {{Takada}}, \citenamefont
  {{Diemer}}, \citenamefont {{Kravtsov}}, \citenamefont {{Dalal}},
  \citenamefont {{More}}, \citenamefont {{Murata}}, \citenamefont
  {{Mandelbaum}}, \citenamefont {{Rozo}}, \citenamefont {{Rykoff}},
  \citenamefont {{Oguri}},\ and\ \citenamefont {{Spergel}}}]{More2016}%
  \BibitemOpen
  \bibfield  {author} {\bibinfo {author} {\bibfnamefont {S.}~\bibnamefont
  {{More}}}, \bibinfo {author} {\bibfnamefont {H.}~\bibnamefont {{Miyatake}}},
  \bibinfo {author} {\bibfnamefont {M.}~\bibnamefont {{Takada}}}, \bibinfo
  {author} {\bibfnamefont {B.}~\bibnamefont {{Diemer}}}, \bibinfo {author}
  {\bibfnamefont {A.~V.}\ \bibnamefont {{Kravtsov}}}, \bibinfo {author}
  {\bibfnamefont {N.~K.}\ \bibnamefont {{Dalal}}}, \bibinfo {author}
  {\bibfnamefont {A.}~\bibnamefont {{More}}}, \bibinfo {author} {\bibfnamefont
  {R.}~\bibnamefont {{Murata}}}, \bibinfo {author} {\bibfnamefont
  {R.}~\bibnamefont {{Mandelbaum}}}, \bibinfo {author} {\bibfnamefont
  {E.}~\bibnamefont {{Rozo}}}, \bibinfo {author} {\bibfnamefont {E.~S.}\
  \bibnamefont {{Rykoff}}}, \bibinfo {author} {\bibfnamefont {M.}~\bibnamefont
  {{Oguri}}}, \ and\ \bibinfo {author} {\bibfnamefont {D.~N.}\ \bibnamefont
  {{Spergel}}},\ }\href@noop {} {\bibfield  {journal} {\bibinfo  {journal}
  {ArXiv e-prints}\ } (\bibinfo {year} {2016})},\ \Eprint
  {http://arxiv.org/abs/1601.06063} {arXiv:1601.06063} \BibitemShut {NoStop}%
\bibitem [{\citenamefont {{Rykoff}}\ \emph {et~al.}(2014)\citenamefont
  {{Rykoff}}, \citenamefont {{Rozo}}, \citenamefont {{Busha}}, \citenamefont
  {{Cunha}}, \citenamefont {{Finoguenov}}, \citenamefont {{Evrard}},
  \citenamefont {{Hao}}, \citenamefont {{Koester}}, \citenamefont
  {{Leauthaud}}, \citenamefont {{Nord}}, \citenamefont {{Pierre}},
  \citenamefont {{Reddick}}, \citenamefont {{Sadibekova}}, \citenamefont
  {{Sheldon}},\ and\ \citenamefont {{Wechsler}}}]{Rykoff2014}%
  \BibitemOpen
  \bibfield  {author} {\bibinfo {author} {\bibfnamefont {E.~S.}\ \bibnamefont
  {{Rykoff}}}, \bibinfo {author} {\bibfnamefont {E.}~\bibnamefont {{Rozo}}},
  \bibinfo {author} {\bibfnamefont {M.~T.}\ \bibnamefont {{Busha}}}, \bibinfo
  {author} {\bibfnamefont {C.~E.}\ \bibnamefont {{Cunha}}}, \bibinfo {author}
  {\bibfnamefont {A.}~\bibnamefont {{Finoguenov}}}, \bibinfo {author}
  {\bibfnamefont {A.}~\bibnamefont {{Evrard}}}, \bibinfo {author}
  {\bibfnamefont {J.}~\bibnamefont {{Hao}}}, \bibinfo {author} {\bibfnamefont
  {B.~P.}\ \bibnamefont {{Koester}}}, \bibinfo {author} {\bibfnamefont
  {A.}~\bibnamefont {{Leauthaud}}}, \bibinfo {author} {\bibfnamefont
  {B.}~\bibnamefont {{Nord}}}, \bibinfo {author} {\bibfnamefont
  {M.}~\bibnamefont {{Pierre}}}, \bibinfo {author} {\bibfnamefont
  {R.}~\bibnamefont {{Reddick}}}, \bibinfo {author} {\bibfnamefont
  {T.}~\bibnamefont {{Sadibekova}}}, \bibinfo {author} {\bibfnamefont {E.~S.}\
  \bibnamefont {{Sheldon}}}, \ and\ \bibinfo {author} {\bibfnamefont {R.~H.}\
  \bibnamefont {{Wechsler}}},\ }\href {\doibase 10.1088/0004-637X/785/2/104}
  {\bibfield  {journal} {\bibinfo  {journal} {\apj}\ }\textbf {\bibinfo
  {volume} {785}},\ \bibinfo {eid} {104} (\bibinfo {year} {2014})},\ \Eprint
  {http://arxiv.org/abs/1303.3562} {arXiv:1303.3562} \BibitemShut {NoStop}%
\bibitem [{\citenamefont {{Rozo}}\ \emph {et~al.}(2015)\citenamefont {{Rozo}},
  \citenamefont {{Rykoff}}, \citenamefont {{Becker}}, \citenamefont
  {{Reddick}},\ and\ \citenamefont {{Wechsler}}}]{Rozo2015}%
  \BibitemOpen
  \bibfield  {author} {\bibinfo {author} {\bibfnamefont {E.}~\bibnamefont
  {{Rozo}}}, \bibinfo {author} {\bibfnamefont {E.~S.}\ \bibnamefont
  {{Rykoff}}}, \bibinfo {author} {\bibfnamefont {M.}~\bibnamefont {{Becker}}},
  \bibinfo {author} {\bibfnamefont {R.~M.}\ \bibnamefont {{Reddick}}}, \ and\
  \bibinfo {author} {\bibfnamefont {R.~H.}\ \bibnamefont {{Wechsler}}},\ }\href
  {\doibase 10.1093/mnras/stv1560} {\bibfield  {journal} {\bibinfo  {journal}
  {\mnras}\ }\textbf {\bibinfo {volume} {453}},\ \bibinfo {pages} {38}
  (\bibinfo {year} {2015})},\ \Eprint {http://arxiv.org/abs/1410.1193}
  {arXiv:1410.1193} \BibitemShut {NoStop}%
\bibitem [{\citenamefont {{Simet}}\ \emph {et~al.}(2016)\citenamefont
  {{Simet}}, \citenamefont {{McClintock}}, \citenamefont {{Mandelbaum}},
  \citenamefont {{Rozo}}, \citenamefont {{Rykoff}}, \citenamefont {{Sheldon}},\
  and\ \citenamefont {{Wechsler}}}]{Simet2016}%
  \BibitemOpen
  \bibfield  {author} {\bibinfo {author} {\bibfnamefont {M.}~\bibnamefont
  {{Simet}}}, \bibinfo {author} {\bibfnamefont {T.}~\bibnamefont
  {{McClintock}}}, \bibinfo {author} {\bibfnamefont {R.}~\bibnamefont
  {{Mandelbaum}}}, \bibinfo {author} {\bibfnamefont {E.}~\bibnamefont
  {{Rozo}}}, \bibinfo {author} {\bibfnamefont {E.}~\bibnamefont {{Rykoff}}},
  \bibinfo {author} {\bibfnamefont {E.}~\bibnamefont {{Sheldon}}}, \ and\
  \bibinfo {author} {\bibfnamefont {R.~H.}\ \bibnamefont {{Wechsler}}},\
  }\href@noop {} {\bibfield  {journal} {\bibinfo  {journal} {ArXiv e-prints}\ }
  (\bibinfo {year} {2016})},\ \Eprint {http://arxiv.org/abs/1603.06953}
  {arXiv:1603.06953} \BibitemShut {NoStop}%
\bibitem [{\citenamefont {{Farahi}}\ \emph {et~al.}(2016)\citenamefont
  {{Farahi}}, \citenamefont {{Evrard}}, \citenamefont {{Rozo}}, \citenamefont
  {{Rykoff}},\ and\ \citenamefont {{Wechsler}}}]{Farahi2016}%
  \BibitemOpen
  \bibfield  {author} {\bibinfo {author} {\bibfnamefont {A.}~\bibnamefont
  {{Farahi}}}, \bibinfo {author} {\bibfnamefont {A.~E.}\ \bibnamefont
  {{Evrard}}}, \bibinfo {author} {\bibfnamefont {E.}~\bibnamefont {{Rozo}}},
  \bibinfo {author} {\bibfnamefont {E.~S.}\ \bibnamefont {{Rykoff}}}, \ and\
  \bibinfo {author} {\bibfnamefont {R.~H.}\ \bibnamefont {{Wechsler}}},\
  }\href@noop {} {\bibfield  {journal} {\bibinfo  {journal} {ArXiv e-prints}\ }
  (\bibinfo {year} {2016})},\ \Eprint {http://arxiv.org/abs/1601.05773}
  {arXiv:1601.05773} \BibitemShut {NoStop}%
\bibitem [{\citenamefont {{Nipoti}}\ \emph {et~al.}(2008)\citenamefont
  {{Nipoti}}, \citenamefont {{Ciotti}}, \citenamefont {{Binney}},\ and\
  \citenamefont {{Londrillo}}}]{Nipoti2008}%
  \BibitemOpen
  \bibfield  {author} {\bibinfo {author} {\bibfnamefont {C.}~\bibnamefont
  {{Nipoti}}}, \bibinfo {author} {\bibfnamefont {L.}~\bibnamefont {{Ciotti}}},
  \bibinfo {author} {\bibfnamefont {J.}~\bibnamefont {{Binney}}}, \ and\
  \bibinfo {author} {\bibfnamefont {P.}~\bibnamefont {{Londrillo}}},\ }\href
  {\doibase 10.1111/j.1365-2966.2008.13192.x} {\bibfield  {journal} {\bibinfo
  {journal} {\mnras}\ }\textbf {\bibinfo {volume} {386}},\ \bibinfo {pages}
  {2194} (\bibinfo {year} {2008})},\ \Eprint {http://arxiv.org/abs/0802.1122}
  {arXiv:0802.1122} \BibitemShut {NoStop}%
\bibitem [{\citenamefont {{Berezhiani}}\ and\ \citenamefont
  {{Khoury}}(2015)}]{Berezhiani2015}%
  \BibitemOpen
  \bibfield  {author} {\bibinfo {author} {\bibfnamefont {L.}~\bibnamefont
  {{Berezhiani}}}\ and\ \bibinfo {author} {\bibfnamefont {J.}~\bibnamefont
  {{Khoury}}},\ }\href@noop {} {\bibfield  {journal} {\bibinfo  {journal}
  {ArXiv e-prints}\ } (\bibinfo {year} {2015})},\ \Eprint
  {http://arxiv.org/abs/1507.01019} {arXiv:1507.01019} \BibitemShut {NoStop}%
\bibitem [{\citenamefont {{Kahlhoefer}}\ \emph {et~al.}(2014)\citenamefont
  {{Kahlhoefer}}, \citenamefont {{Schmidt-Hoberg}}, \citenamefont
  {{Frandsen}},\ and\ \citenamefont {{Sarkar}}}]{Kahlhoefer2014}%
  \BibitemOpen
  \bibfield  {author} {\bibinfo {author} {\bibfnamefont {F.}~\bibnamefont
  {{Kahlhoefer}}}, \bibinfo {author} {\bibfnamefont {K.}~\bibnamefont
  {{Schmidt-Hoberg}}}, \bibinfo {author} {\bibfnamefont {M.~T.}\ \bibnamefont
  {{Frandsen}}}, \ and\ \bibinfo {author} {\bibfnamefont {S.}~\bibnamefont
  {{Sarkar}}},\ }\href {\doibase 10.1093/mnras/stt2097} {\bibfield  {journal}
  {\bibinfo  {journal} {\mnras}\ }\textbf {\bibinfo {volume} {437}},\ \bibinfo
  {pages} {2865} (\bibinfo {year} {2014})},\ \Eprint
  {http://arxiv.org/abs/1308.3419} {arXiv:1308.3419 [astro-ph.CO]} \BibitemShut
  {NoStop}%
\end{thebibliography}%
\end{document}